# Thermodynamical aspects of Bianchi type-I Universe in quadratic form of $f(Q)$ gravity and observational constraints

M. Koussour,[1, *] S. H. Shekh,[2, †] M. Govender,[3, ‡] and M. Bennai[1, 4, §]

[1]*Quantum Physics and Magnetism Team, LPMC, Faculty of Science Ben M'sik,*
*Casablanca Hassan II University, Morocco.*
[2]*Department of Mathematics. S. P. M. Science and Gilani Arts Commerce College,*
*Ghatanji, Dist. Yavatmal, Maharashtra-445301, India.*
[3]*Department of Mathematics, Steve Biko Campus, Durban University of Technology, Durban 4000, South Africa.*
[4]*Lab of High Energy Physics, Modeling and Simulations, Faculty of Science,*
*University Mohammed V-Agdal, Rabat, Morocco.*

(Dated: November 28, 2022)

In this paper, we discuss the Bianchi type-I cosmological model in the framework of symmetric teleparallel gravity say $f(Q)$ gravity in which the non-metricity term $Q$ is responsible for the gravitational interaction. We consider a special form of the $f(Q)$ function which can be cast as $f(Q) = \lambda Q^n$, where $\lambda$ and $n$ both are the dynamical model parameters. Such a choice can be viewed as a hybrid scale factor that leads to a relation between cosmic time and redshift as $t = \left(\frac{\alpha t_0}{\beta}\right) W\left[\frac{\beta}{\alpha}e^{\frac{\beta - \ln(1+z)}{\alpha}}\right]$ which describes a $\Lambda$CDM model of the Universe with the expansion evolving from decelerating to an acceleration phase. The best values for the model parameters i.e. $\alpha$ and $\beta$ that would accord with the most current observational datasets are then estimated. We make use of 57 points from the Hubble dataset, 1048 points from the supernovae of type Ia dataset and 6 points from the BAO dataset. We use the Markov Chain Monte Carlo (MCMC) technique in conjunction with Bayesian analysis and the likelihood function. Further, we study the validity of our model with the investigation of the thermodynamical quantities, energy conditions along with some physical variables such as the EoS, and jerk parameters. Next, our results are discussed in light of current observational data and trends.

## I. INTRODUCTION

Contemporary astrophysical observations demonstrate that the present-day cosmos experiences an accelerated expansion due to almost mystical energy with a large negative pressure called Dark Energy (DE) [1, 2] along with other observations such as Cosmic Microwave Background (CMB) anisotropies measured with Wilkinson Microwave Anisotropy Probe (WMAP) satellite [3] and Large-Scale Structure (LSS) [4] advocated just about two-thirds of the Universe be made up of DE while the remaining consists of relativistic Dark Matter (DM) and baryons [5]. The ultimate nature of DE is commonly categorized by the ratio of spatially homogeneous pressure to the energy density of DE, the so-called Equation of State (EoS) parameter. According to recent cosmological observations, one can understand the worth of the EoS parameter, $\omega < -1/3$ is a prerequisite for the speeded-up cosmic expansion. The crucial candidates in this classification are scalar field models in which the EoS parameter is $-1 < \omega < -1/3$ called as a Quintessence field DE [6, 7] whereas $\omega < -1$ as phantom field DE [8]. Amongst this, the phantom field DE has attracted more attention in line with its strange properties. The phantom model characterizes to devlop DE that perpetuates in an exciting future spreading, leading to finite-time future singularity. It also violates the Strong Energy Conditions (SEC) that help to constrain wormholes [9]. Further, the present value of the EoS parameter for DE attained by the combined observations of WMAP9 [10] and the $H_0$ measurements, Supernovae of type Ia (SNe Ia), CMB, and BAO (Baryon Acoustic Oscillations) show that $\omega_0 = -1.084 \pm 0.063$. This is because, in 2015, the Planck collaboration showed that $\omega_0 = -1.006 \pm 0.0451$ [11] and in 2018 we had a refinement of $\omega_0 = -1.028 \pm 0.032$ [12].

To solve the problem of the nature of DE, many other alternatives have been proposed. Motivated by research on black hole thermodynamics [13], Hooft proposed for the first time the famous holographic principle [14]. Later on, the holographic principle was applied to the DE problem, pramoting a new model of DE referred to as holographic DE in which the energy density relies on physical quantities on the boundary of the Universe these are the reduced Planck mass and the cosmological length scale which is chosen as the future event horizon of the Universe [15]. Several researchers [16, 17] have inspected altered holographic DE by making an allowance for different types of cut-offs, for example, Hubble's and Granda-Oliveros. In the current analysis, we investigated the DE propagating from the gravitational region as an alternative to the matter part. Such a line of attack has been extensively used in the collected works in the framework of so-called modified theories of gravity [18–20].

[*] pr.mouhssine@gmail.com
[†] da_salim@rediff.com
[‡] megandhreng@dut.ac.za
[§] mdbennai@yahoo.fr

The fundamentals in the General Relativity (GR) is the curvature imported from Riemannian geometry which is described by the Ricci scalar $R$. The modified $f(R)$ gravity is a simple modification of GR, replacing the Ricci scalar $R$ with some general function of $R$ [18–20]. Moreover, other alternatives to GR such as the teleparallel equivalent of GR (TEGR), in which the gravitational interactions are outlined by the concept of torsion $T$. In GR, the Levi-Civita connection is associated with curvature, but zero torsion, while in teleparallelism, the Weitzenbock connection is associated with torsion, but zero curvature [21]. In the same way, the $f(T)$ gravity is the simplest modification of TEGR. Recently, a new theory of gravity has been proposed called the symmetric teleparallel equivalent of GR (STEGR), in which the gravitational interactions are outlined by the concept of non-metricity $Q$ with zero torsion and curvature [22]. The non-metricity in Weyl geometry (generalization of Riemannian geometry) represents the variation of a vector length in parallel transport. In Weyl geometry, the covariant derivative of the metric tensor is not equal to zero but is determined mathematically by the non-metricity tensor i.e. $Q_{\gamma\mu\nu} = \nabla_\gamma g_{\mu\nu}$. In addition, the so-called $f(Q)$ gravity is the simplest modification of STEGR. The first cosmological solutions in $f(Q)$ gravity appear in references [23, 24], while $f(Q)$ cosmography with energy conditions can be found in [25, 26]. The cosmological clarifications and evolution index of matter perturbations have been examined for a polynomial functional form of $f(Q)$ gravity in [27]. Harko et al. [28] investigated the coupling matter in $f(Q)$ gravity by supposing a power-law function. In a recent exposition of holographic dark energy models in $f(Q)$ gravity, Shekh [29] investigated Holographic and Renyi holographic dark energy inflation with Hubble and Granda-Oliveros cut-off and observed that the models are ruled out by Galaxy Clustering Statistics with range $-1.33 \leq \omega \leq -0.79$. Furthermore, the Universe exhibits the transition from the early deceleration phase to the current acceleration phase. Koussour et al. [30] studied a spatially homogeneous and isotropic FLRW cosmological model in the framework of symmetric teleparallel gravity of the form $f(Q) = \alpha Q^n + \beta$ with Lambert function. This model achieved quintessence field in present epoch of the Universe and $\Lambda$CDM model in future with the strong justification of jerk and statefinder parameters.

Diverse research has shown that the geometry of the Universe at the end of the inflationary epoch is homogeneous and isotropic [31], with FLRW models playing an essential role in this phase. But the appearance of anomalies in CMB owing to quantum fluctuations of the inflation phase validates the existence of an anisotropic stage which then transformed into an isotropic stage. Newly, with the onset of Planck observational data [12], Bianchi's cosmological models outlining the anisotropic Universe have attracted to many authors. In the literatures, there are various types of anisotropic and inhomogeneous Bianchi space-times. The Bianchi type-I space-time is the simplest mathematical study which describe an anisotropic and homogeneous Universe. Furthermore, it is defined as a straight generalization of the FLRW Universe with various scale factors in its spatial direction. Various Bianchi type-I cosmological models have been studied by so many researchers like [32–34]. Recently, to examine the early evolution of the Universe, De et al. [35] obtained the Bianchi-I field equations in $f(Q)$ gravity and some cosmological parameters are regarded in this background.

Motivated by previous work and results, we analyze $f(Q)$ gravity in the anisotropic background and discuss some physical parameters along with energy conditions, thermodynamical aspects like temperature and entropy density, and jerk parameter of the Universe. Then we calculate the model parameters that are most likely to accord with the most current observational datasets like 57 points from the Hubble dataset, 1048 points from the SNe Ia dataset, and 6 points from the BAO dataset. Along with the Markov Chain Monte Carlo (MCMC) approach is used for Bayesian analysis and the likelihood function.

The present paper is organized as follows: In Sec. II we present the $f(Q)$ gravity formalism and associated field equation in view of the Bianchi type-I space-time. In Sec. III we derive the solution of the field equations by using some physical conditions with the help of the hybrid expansion law. Using 57 points from the Hubble dataset, 1048 points from the SNe Ia dataset and, 6 points from the BAO dataset, we constrained the model parameters in Sec. IV. In Sec. V we discuss some physical and kinematical properties of the derived model. Further, in Secs. VI, VII and VIII we analyze the behavior of the energy conditions, thermodynamical aspects, and jerk parameter, respectively. Finally, in Sec. IX we discuss the results by concluding remarks.

## II. $f(Q)$ GRAVITY FORMALISM AND FIELD EQUATIONS

The action of $f(Q)$ symmetric teleparallel gravity has been well presented in the current literature. For a more detailed discussion the reader is referred to the works contained in references [23–25]. Starting off with the action

$$S = \int \left[\frac{1}{2\kappa}f(Q) + L_m\right] d^4x\sqrt{-g}, \quad (1)$$

where $f(Q)$ is an arbitrary function of the non-metricity scalar $Q$, $g$ is the determinant of the metric tensor $g_{\mu\nu}$ and $L_m$ is the matter Lagrangian density.

The non-metricity scalar $Q$ can be obtained by the disformation tensor $L^{\gamma}{}_{\mu\nu}$ as

$$Q \equiv -g^{\mu\nu}(L^{\beta}{}_{\alpha\mu}L^{\alpha}{}_{\nu\beta} - L^{\beta}{}_{\alpha\beta}L^{\alpha}{}_{\mu\nu}), \quad (2)$$

where the disformation tensor $L^{\gamma}{}_{\mu\nu}$ is defined by

$$L^{\beta}{}_{\alpha\gamma} = -\frac{1}{2}g^{\beta\eta}(\nabla_{\gamma}g_{\alpha\eta} + \nabla_{\alpha}g_{\eta\gamma} - \nabla_{\eta}g_{\alpha\gamma}). \quad (3)$$

The non-metricity tensor and its traces can be formulated as

$$Q_{\beta\mu\nu} = \nabla_{\beta}g_{\mu\nu}, \quad (4)$$

$$Q_{\beta} = g^{\mu\nu}Q_{\beta\mu\nu} \quad \widetilde{Q}_{\beta} = g^{\mu\nu}Q_{\mu\beta\nu}. \quad (5)$$

In addition, we define the superpotential (the non-metricity conjugate) tensor as

$$P^{\beta}{}_{\mu\nu} = -\frac{1}{2}L^{\beta}{}_{\mu\nu} + \frac{1}{4}(Q^{\beta} - \widetilde{Q}^{\beta})g_{\mu\nu} - \frac{1}{4}\delta^{\beta}_{(\mu}Q_{\nu)}.$$

Using the above definition, the non-metricity scalar is given as

$$Q = -Q_{\beta\mu\nu}P^{\beta\mu\nu}. \quad (6)$$

Furthermore, the energy-momentum tensor for the matter is defined by

$$T_{\mu\nu} = \frac{-2}{\sqrt{-g}}\frac{\delta\left(\sqrt{-g}L_m\right)}{\delta g^{\mu\nu}}. \quad (7)$$

Now, the variation of the action ($S$) in Eq. (1) with respect to the metric tensor $g_{\mu\nu}$, yields the following gravitational field equations

$$\frac{2}{\sqrt{-g}}\nabla_{\beta}(\sqrt{-g}f_Q P^{\beta}{}_{\mu\nu}) - \frac{1}{2}fg_{\mu\nu} + f_Q(P_{\nu\rho\sigma}Q_{\mu}{}^{\rho\sigma} - 2P_{\rho\sigma\mu}Q^{\rho\sigma}{}_{\nu}) = \kappa T_{\mu\nu}, \quad (8)$$

where $f_Q = \frac{df}{dQ}$ and $\nabla_{\beta}$ denotes the covariant derivative.

Taking into account the spatially homogeneous and anisotropic Universe, we assume the following flat Bianchi type-I metric for our analysis

$$ds^2 = -dt^2 + A^2(t)dx^2 + B^2(t)\left(dy^2 + dz^2\right), \quad (9)$$

where the metric potentials $A(t)$ and $B(t)$ are functions of cosmic time $t$ only. The corresponding non-metricity scalar is given by [35]

$$Q = -2\left(\frac{\dot{B}}{B}\right)^2 - 4\frac{\dot{A}}{A}\frac{\dot{B}}{B}. \quad (10)$$

The energy-momentum tensor for the anisotropic perfect fluid is defined as

$$T^{\mu}_{\nu} = diag\left[-\rho, p_x, p_y, p_z\right], \quad (11)$$

where $\rho$ is the energy density of the Universe, $p_x$, $p_y$ and $p_z$ are pressures in the directions of $x$, $y$ and $z$ axes respectively. Considering the anisotropy in pressure and the equation of state (EoS) parameter, we have

$$T^{\mu}_{\nu} = \text{diag}(-1, \omega_x, \omega_y, \omega_z)\rho, \quad (12)$$
$$= \text{diag}(-1, \omega, (\omega + \delta), (\omega + \delta))\rho,$$

where $\delta$ is the skewness parameter, i.e. the deviation from $\omega$ along the $y$ and $z$ axes ($\omega_x = \omega$). The parameters $\omega$ and $\delta$ are not constants and may be functions of cosmic time $t$. Using a co-moving coordinate system, the field equations (8) for the metric given in Eq. (9) with the help of Eq. (12) take the form [35]

$$\frac{f}{2} + f_Q \left[ 4\frac{\dot{A}}{A}\frac{\dot{B}}{B} + 2\left(\frac{\dot{B}}{B}\right)^2 \right] = \rho, \tag{13}$$

$$\frac{f}{2} - f_Q \left[ -2\frac{\dot{A}}{A}\frac{\dot{B}}{B} - 2\frac{\ddot{B}}{B} - 2\left(\frac{\dot{B}}{B}\right)^2 \right] + 2\frac{\dot{B}}{B}\dot{Q}f_{QQ} = -\omega\rho, \tag{14}$$

$$\frac{f}{2} - f_Q \left[ -3\frac{\dot{A}}{A}\frac{\dot{B}}{B} - \frac{\ddot{A}}{A} - \frac{\ddot{B}}{B} - \left(\frac{\dot{B}}{B}\right)^2 \right] + \left(\frac{\dot{A}}{A} + \frac{\dot{B}}{B}\right)\dot{Q}f_{QQ} = -(\omega + \delta)\rho \tag{15}$$

where the dot $(\cdot)$ denotes the derivative with respect to cosmic time $t$.

The spatial volume for the Bianchi type-I model is given by

$$V = a^3(t) = AB^2. \tag{16}$$

where $a(t)$ is the scale factor of the Universe.

The deceleration parameter $(q)$ is defined as

$$q = \frac{d}{dt}\left(\frac{1}{H}\right) - 1. \tag{17}$$

where $H$ is the Hubble parameter. The deceleration parameter is the quantity that describes the evolution of the expansion of the Universe. This parameter is positive $(q > 0)$ when the Universe is decelerating over time and is negative $(q < 0)$ in the case of an accelerating Universe. The average Hubble parameter $H$ is given by

$$H = \frac{1}{3}\left(H_x + H_y + H_z\right), \tag{18}$$

where $H_x$, $H_y$ and $H_z$ are the directional Hubble parameters along $x$, $y$ and $z$ axes respectively. In view of Eq. (9), these parameters are takes the form $H_x = \frac{\dot{A}}{A}$ and $H_y = H_z = \frac{\dot{B}}{B}$.

The Hubble parameter, spatial volume, and scale factor are connected by

$$H = \frac{1}{3}\frac{\dot{V}}{V} = \frac{1}{3}\left[\frac{\dot{A}}{A} + 2\frac{\dot{B}}{B}\right] = \frac{\dot{a}}{a}. \tag{19}$$

The scalar expansion $\theta(t)$, shear scalar $\sigma^2(t)$ and the mean anisotropic parameters $\Delta$ are given by

$$\theta(t) = \frac{\dot{A}}{A} + 2\frac{\dot{B}}{B}, \tag{20}$$

$$\sigma^2(t) = \frac{1}{3}\left(\frac{\dot{A}}{A} - \frac{\dot{B}}{B}\right)^2, \tag{21}$$

$$\Delta = \frac{1}{3}\sum_{i=1}^{3}\left(\frac{H_i - H}{H}\right)^2, \tag{22}$$

where $H_i$, $i = 1, 2, 3$ are directional Hubble parameters.

### III. SOLUTIONS OF THE FIELD EQUATIONS

The field equations (13)-(15) constitute a system of three independent equations in seven unknowns, $A$, $B$, $f(Q)$, $\rho$, $\omega$, $\delta$, and $Q$. So the system is initially undetermined. To find exact solutions of the field equations, additional physical constraints are needed. We first propose a physical condition that the shear is proportional to the expansion scalar $\left(\sigma^2 \propto \theta\right)$ and this leads to the relation

$$A = B^m, \tag{23}$$

where $m \neq 1$ is an arbitrary constant. For the case $m = 1$ we obatin an isotropic model, otherwise the model is anisotropic. The main reasons behind the assumptions that led to this condition are discussed in detail here [35]. Observations of the velocity redshift relation for extragalactic sources indicate that the Hubble expansion of the Universe can reach isotropy when $\frac{\sigma^2}{\theta}$ is constant. This condition has been used in many studies. In addition, we consider the following functional form as,

$$f(Q) = \lambda Q^n, \tag{24}$$

where $\lambda$ is an arbitrary constant. If we consider the case where $n = 2$, then $f(Q) = \lambda Q^2$, we get the ordinary field equations of quadratic $f(Q)$ theory of gravity governing the Bianchi type-I Universe. Finally, we add a constraint to the scale factor which takes the form of the hybrid expansion law (HEL), this form produces a transition of the Universe from deceleration to the acceleration phase and it is given as,

$$a(t) = a_0 \left(\frac{t}{t_0}\right)^\alpha e^{\beta\left(\frac{t}{t_0}-1\right)}, \tag{25}$$

where $\alpha \geq 0$, and $\beta \geq 0$ are constants and $a_0$ and $t_0$ denote the current scale factor value and age of the Universe, respectively. It is known in the literature that this law is a generalization of both the power-law cosmology and the exponential law cosmology. For $\alpha = 0$ hybrid expansion law in Eq. (25) reduces to exponential law $a = a_0 e^{\beta\left(\frac{t}{t_0}-1\right)}$ and for $\beta = 0$, it reduces to power-law $a = a_0 \left(\frac{t}{t_0}\right)^\alpha$. Using the same law of hybrid expansion explored by Shekh et al. [36] in which they showed that the holographic dark energy fluid with perfect fluid mimicked a $\Lambda$CDM model. By transforming the above hybrid law into regular bounce the same author [37] discussed the dynamical analysis with thermodynamic aspects of anisotropic DE bounce cosmological model in $f(R, G)$ gravity and observed that EoS parameter takes a negative value which is an acceptable form observed by SNIa observations.

Now, from Eqs. (16), (23) and (25) we obtain the metric potentials as

$$A(t) = \left[a_0 \left(\frac{t}{t_0}\right)^\alpha e^{\beta\left(\frac{t}{t_0}-1\right)}\right]^{\frac{3m}{2+m}}, \tag{26}$$

$$B(t) = \left[a_0 \left(\frac{t}{t_0}\right)^\alpha e^{\beta\left(\frac{t}{t_0}-1\right)}\right]^{\frac{3}{2+m}}. \tag{27}$$

Through the above equations, it is clear that the metric potentials initially vanish. Hence, the model has an early singularity. Further, the scale factors diverge to infinity when $t \to \infty$. Consequently, there will be Big Rip at minimum in the distant future because the metric potentials tend to infinity at $t \to \infty$. Using Eqs. (26) and (27), the metric in Eq. (9) becomes,

$$ds^2 = -dt^2 + \left[a_0 \left(\frac{t}{t_0}\right)^\alpha e^{\beta\left(\frac{t}{t_0}-1\right)}\right]^{\frac{6m}{2+m}} dx^2 + \left[a_0 \left(\frac{t}{t_0}\right)^\alpha e^{\beta\left(\frac{t}{t_0}-1\right)}\right]^{\frac{6}{2+m}} \left(dy^2 + dz^2\right). \tag{28}$$

The mean Hubble parameter is given by

$$H = \frac{\alpha}{t} + \frac{\beta}{t_0}. \tag{29}$$

The redshift parameter $z$ for our model is determined as

$$z = -1 + \frac{a_0}{a} = -1 + \frac{1}{a_0 \left(\frac{t}{t_0}\right)^\alpha e^{\beta\left(\frac{t}{t_0}-1\right)}}. \tag{30}$$

where $a_0 = 1$ is the present value of the scale factor.

Using the above relation, we obtain

$$\frac{1}{(1+z)} = a_0 \left(\frac{t}{t_0}\right)^\alpha e^{\beta\left(\frac{t}{t_0}-1\right)}. \tag{31}$$

After simple calculations, we find

$$\frac{\beta}{\alpha} e^{\frac{\beta-\ln(1+z)}{\alpha}} = \frac{\beta}{\alpha}\left(\frac{t}{t_0}\right) e^{\frac{\beta}{\alpha}\left(\frac{t}{t_0}\right)}. \tag{32}$$

By inserting Lambert's function ($W$) on both sides, we get

$$W\left[\frac{\beta}{\alpha} e^{\frac{\beta-\ln(1+z)}{\alpha}}\right] = \frac{\beta}{\alpha}\left(\frac{t}{t_0}\right), \tag{33}$$

where $W\left[\frac{\beta}{\alpha}\left(\frac{t}{t_0}\right) e^{\frac{\beta}{\alpha}\left(\frac{t}{t_0}\right)}\right] = \frac{\beta}{\alpha}\left(\frac{t}{t_0}\right)$.

Finally, the redshift–time relation $t(z)$ can be expressed as

$$t(z) = \left(\frac{\alpha t_0}{\beta}\right) g(z), \quad \text{where} \quad g(z) = W\left[\frac{\beta}{\alpha} e^{\frac{\beta-\ln(1+z)}{\alpha}}\right]. \tag{34}$$

Using Eq. (29), the Hubble parameter expression in terms of redshift may be found as

$$H(z) = \frac{H_0 \beta}{\alpha + \beta}\left[\frac{1}{g(z)} + 1\right] \tag{35}$$

## IV. COSMOLOGICAL CONSTRAINTS

The cosmological constraints of the under consideration model are presented in this section. We are able to constrain the parameters such as $\alpha$ and $\beta$ thanks to the statistical technique we employ. With the usual Bayesian technique, we decided on the Markov Chain Monte Carlo (MCMC). The datasets below are utilized:

### A. $H(z)$ dataset:

Sharov and Vasiliev [38] have created a catalog of 57 Hubble ($H(z)$) parameter data points in the redshift range $0.07 \leq z \leq 2.41$. This $Hz$ dataset was calculated using differential galaxy ages $\triangle t$ [39–42] and line-of-sight BAO data [43–47]. The Ref. [38] displays a comprehensive list of datasets. We employed the Chi-square test using MCMC simulation to estimate the model parameters $\alpha$ and $\beta$. The formula for the Chi-square ($\chi^2$) function is,

$$\chi^2_{Hz}(\alpha, \beta) = \sum_{i=1}^{57} \frac{\left[H_{th}(z_i, \alpha, \beta) - H_{obs}\right]^2}{\sigma(z_i)^2}, \quad (36)$$

where $H_{obs}$ denotes the Hubble parameter values that have been observed, $H_{th}(z_i, \alpha, \beta)$ denotes the Hubble parameter when the model parameters are used, and $\sigma(z_j)$ denotes the standard deviation.

### B. SNe dataset:

As the fact of expansion of universe which is supported by the SNe Ia observation. Significantly, the SNe Ia data is recorded from the Panoramic Survey Telescope and Rapid Response system (Pan-STARSS1), Sloan Digital Sky Survey (SDSS), Supernova Legacy Survey (SNLS), and Hubble Space Telescope (HST) survey [48]. With the use of Pantheon sample data set of 1048 points of distance moduli $\mu_j$ towards the range $0.01 < z_j < 2.26$ for various redshifts. We observe the analysis using the expression of $\chi^2_{SNe}$ as

$$\chi^2_{SNe}(\alpha, \beta) = \sum_{i,j=1}^{1048} \nabla \mu_i (C_{SNe}^{-1})_{ij} \nabla \mu_j, \quad (37)$$

where $\nabla \mu_i = \mu^{th}(z_i, \alpha, \beta) - \mu_i^{obs}$, $C_{SNe}$ is the covariance metric, and $\mu$ denotes the distance moduli is given by,

$$\mu(z) = 5 \log_{10} D_L(z) + \mu_0, \quad (38)$$

with

$$\mu_0 = 5 \log \left(\frac{H_0^{-1}}{Mpc}\right) + 25, \quad (39)$$

and

$$D_L(z) = (1+z) \int_0^z \frac{c \, dz'}{H(z')}. \quad (40)$$

here, $c$ is the speed of light.

### C. BAO dataset:

The BAO distance dataset consists of BAO measurements at six distinct redshifts for the 6dFGS, SDSS, and WiggleZ surveys. The sound horizon $r_s$ at the photon decoupling epoch $z_*$, which is defined by the following relation, determines the characteristic scale of BAO:

$$r_s(z_*) = \frac{c}{\sqrt{3}} \int_0^{\frac{1}{1+z_*}} \frac{da}{a^2 H(a) \sqrt{1 + (3\Omega_{b0}/4\Omega_{\gamma 0})a}}. \quad (41)$$

The current density of baryons and photons are indicated here by the symbols $\Omega_{b0}$ and $\Omega_{\gamma 0}$, respectively. BAO measurements employ the following relations:

$$\triangle \theta = \frac{r_s}{d_A(z)}, \quad (42)$$

$$d_A(z) = \int_0^z \frac{dz'}{H(z')}, $$

$$\triangle z = H(z) r_s. \quad (43)$$

Here, $\triangle z$ is the detected redshift separation of the BAO features in the two - point correlation function of the galaxy distribution on the sky along the line of sight, $d_A$ is the measured angular diameter distance, and $\triangle \theta$ indicates the measured angular separation of the BAO feature. The redshift at the time of photon decoupling is assumed to be $z_* \approx 1091$ in this study, and $d_A(z)$ the co-moving angular diameter distance together with the dilation scale $D_V(z) = \left[d_A(z)^2 z / H(z)\right]^{1/3}$ are used in the BAO dataset of six locations for $d_A(z_*)/D_V(z_{BAO})$ calculations [49–54]. For the BAO dataset, the chi-square function is assume as [54]

$$\chi^2_{BAO} = X^T C_{BAO}^{-1} X, \quad (44)$$

where $X$ based on the survey considered and $C_{BAO}^{-1}$ is the inverse covariance matrix [54].

TABLE I. By employing the Hubble and combined observational dataset, the minimized constraining findings on two model parameters are shown.

| Dataset | $Hz$ dataset | $Hz$+SNe+BAO dataset |
|---|---|---|
| $\alpha$ | $0.477^{+0.047}_{-0.043}$ | $0.559^{+0.026}_{-0.025}$ |
| $\beta$ | $0.476^{+0.033}_{-0.034}$ | $0.432^{+0.027}_{-0.027}$ |

TABLE II. By employing the Hubble and combined observational dataset, the minimized constraining findings on two model parameters are shown.

### D. $H(z)$+SNe+BAO dataset:

The $\chi^2_{tot}$ function for the $Hz$+SNe+BAO dataset is expressed as: $\chi^2_{tot} = \chi^2_{Hz} + \chi^2_{SNe} + \chi^2_{BAO}$. By using MCMC to minimize each $\chi^2$, the constraints on model parameters are determined. Tab. II contains the model parameter values that best suit the data. Figs. 1 and 2 also display the error bar fit for the model under consideration and the $\Lambda$CDM with cosmological constant density parameter $\Omega_{\Lambda_0} = 0.7$, matter density parameter $\Omega_{m_0} = 0.3$ and $H_0 = 69$ km/s/Mpc, respectively. Figs. 3 and 4 display the $1-\sigma$ and $2-\sigma$ contour curves for Hubble and the combined observational dataset.

## V. PHYSICAL AND KINEMATICAL PROPERTIES

In cosmology, there is a set of physical and kinematic parameters whose behavior can generally be studied either by analyzing their expressions or by interpreting the graphic representations. In this section, we will analyze the expressions of some necessary parameters such as mean Hubble parameter $H(t)$, the expansion scalar $\theta(t)$, the shear scalar $\sigma^2(t)$, the mean anisotropic parameter $\Delta$, the spatial volume $V$ and deceleration parameter $q$ defined in the previous section.

The expansion scalar takes the form

$$\theta(t) = 3\left(\frac{\alpha}{t} + \frac{\beta}{t_0}\right). \quad (45)$$

The shear scalar is expressed as

$$\sigma^2(t) = 3\left(\frac{m-1}{m+2}\right)^2\left(\frac{\alpha}{t} + \frac{\beta}{t_0}\right)^2. \quad (46)$$

Using the directional and mean Hubble parameter, the mean anisotropic parameter is obtained as

$$\Delta = \frac{2}{9H^2}\left(H_x - H_y\right)^2 = 2\left(\frac{m-1}{m+2}\right)^2. \quad (47)$$

The spatial volume is derived as

$$V = a_0^3\left(\frac{t}{t_0}\right)^{3\alpha} e^{3\beta\left(\frac{t}{t_0}-1\right)}. \quad (48)$$

The deceleration parameter is derived as

$$q = -1 + \alpha t_0^2 \left(\beta t + \alpha t_0\right)^{-2}. \quad (49)$$

From Eqs. (29)-(48), it is clear that the spatial volume is zero at the early period (i.e. at $t = 0$) and the parameters $H$, $\theta(t)$, and $\sigma^2(t)$ diverge during this epoch, and as $t \to \infty$, spatial volume $V \to \infty$ and $H$, $\theta(t)$, and $\sigma^2(t)$ all tend to zero. Thus, the model initiates evolving with zero volume with an infinite rate of expansion and this expansion rate slows down via the evolution of the studied model. The mean anisotropic parameter is constant (i.e. $\Delta = const \neq 0$) throughout the evolution of the Universe. Hence, the studied model is evolving with an anisotropy. From (49), we observe that $q > 0$ for $t < \frac{t_0}{\beta}\left(\sqrt{\alpha} - \alpha\right)$ and $q < 0$ for $t > \frac{t_0}{\beta}\left(\sqrt{\alpha} - \alpha\right)$. In addition, the transition phase from deceleration to acceleration phase at $t = t_0\left(-\frac{\alpha}{\beta} \pm \frac{\sqrt{\alpha}}{\beta}\right)$ with $0 < \alpha < 1$. Further, we observe that with the negativity of the second term with a negative time indicates an unphysical background of the Big Bang cosmology. We conclude that the cosmic transition may have taken place at $t = \frac{t_0}{\beta}\left(\sqrt{\alpha} - \alpha\right)$. Current observations of SNe Ia, indicate that, at the present day, the Universe is accelerating, and the value of the deceleration parameter is situated in the range of $-1 \leq q < 0$. Using Eq. (49), we plot the behavior of the deceleration parameter in terms of redshift in Fig. 5 for the constrained values of the model parameters $\alpha$ and $\beta$. From Fig. 5, it is clear that our model shows a transition from early deceleration to late-time acceleration which is in keeping with the findings of the work contained in [29, 30] and is in excellent agreement with current observations. According to the Hubble and $Hz$+SNe+BAO datasets, the present value of the deceleration parameter is $q_0 = -0.477$ and $q_0 = -0.427$, respectively.

$$\frac{\sigma^2}{\theta} = \frac{(m-1)^2}{(m+2)^2}\left(\frac{\alpha}{t} + \frac{\beta}{t_0}\right). \quad (50)$$

From Eq. (50), it is clear that $\lim_{t\to\infty}\left(\frac{\sigma^2}{\theta}\right)$ becomes a constant value, this leads to the maintenance of anisotropy in the Universe throughout evolution.

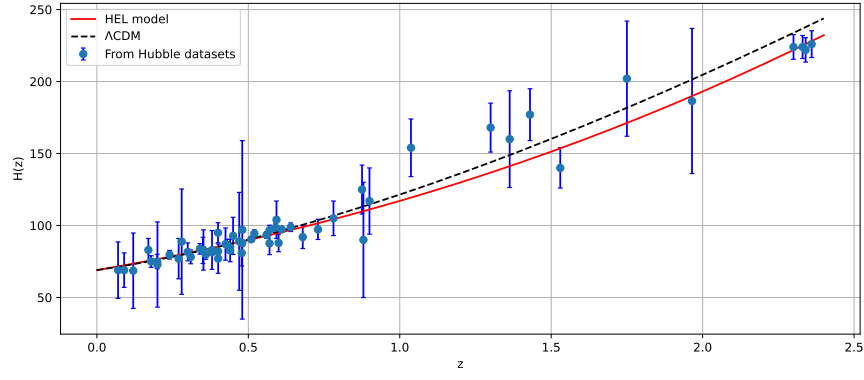

FIG. 1. This graph shows the evolution of $H(z)$ with regard to redshift $z$. The dashed line denotes the "ΛCMD model", while the red line reflects our model. The 57 Hubble dataset with error bar is shown as dots.

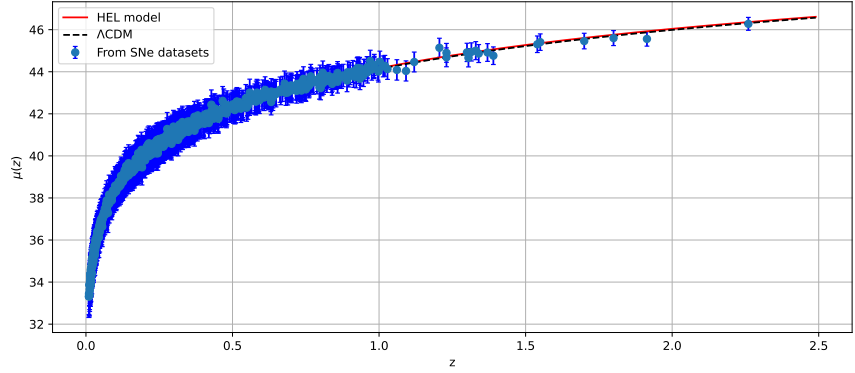

FIG. 2. This graph shows the evolution of $\mu(z)$ with regard to redshift $z$. The dashed line denotes the "ΛCMD model", while the red line reflects our model. The 1048 SNe dataset with error bar is shown as dots.

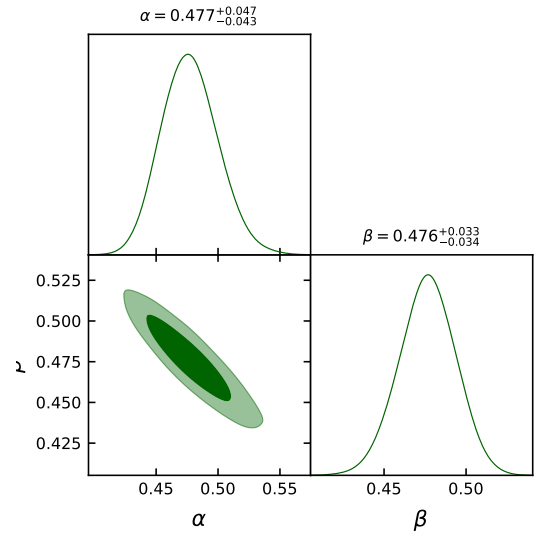

FIG. 3. The Hubble dataset is used to illustrate the minimized constraints on the parameters in the expression of the Hubble parameter $H(z)$ in Eq. (35).

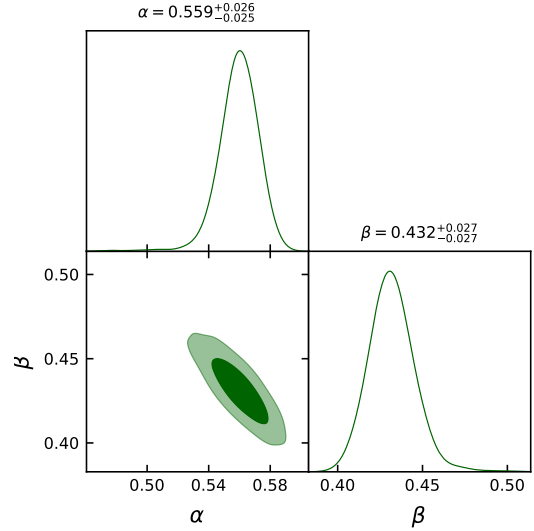

FIG. 4. The combined observational dataset is used to illustrate the minimized constraints on the parameters in the expression of the Hubble parameter $H(z)$ in Eq. (35).

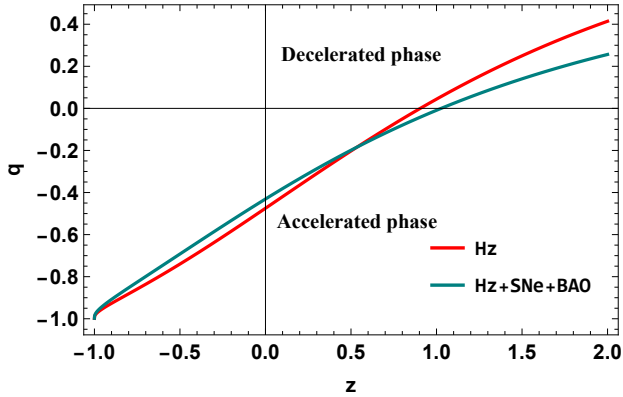

FIG. 5. Evolution of deceleration parameter $q$ in quadratic $f(Q)$ gravity model with the constraints value of the model parameters $\alpha$, $\beta$ versus redshift $z$.

The non-metricity scalar for the model becomes,

$$Q = -\frac{18(2m+1)}{(m+2)^2}\left(\frac{\alpha}{t} + \frac{\beta}{t_0}\right)^2. \quad (51)$$

It is observed that the non-metricity of the Universe for our model is time-dependent. Using (24) and (51) in (13), we obtained the expressions of energy density as

$$\rho = -\frac{486\lambda(2m+1)^2}{(m+2)^4}\left(\frac{\alpha}{t} + \frac{\beta}{t_0}\right)^4. \quad (52)$$

The behavior of energy density for our model versus redshift is shown in Fig. 6. From this figure, it is observed that the energy density is positive and is an increasing function of redshift for all the constrained values of the model parameters from the observation data. Further, when $z = 0$, the energy density of Universe is positive and increases with the value of $z$.

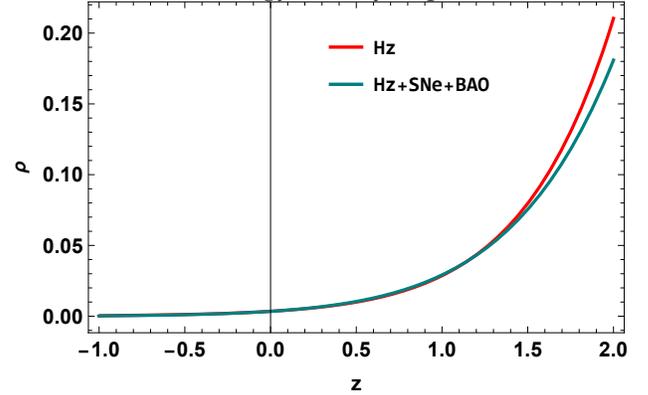

FIG. 6. Evolution of energy density $\rho$ in quadratic $f(Q)$ gravity model with the constraints value of the model parameters $\alpha$, $\beta$ together with $\lambda = -0.5$ and $m = 0.5$ versus redshift $z$.

The anisotropic pressure is given by

$$p = \left[\frac{648\lambda(1+2m)}{(m+2)^3} - \frac{162\lambda(1+2m)^2}{(m+2)^4}\right]\left(\frac{\alpha}{t} + \frac{\beta}{t_0}\right)^4 - \frac{648\alpha\lambda(1+2m)}{t^2(m+2)^3}\left(\frac{\alpha}{t} + \frac{\beta}{t_0}\right)^2 \quad (53)$$

rameters is obtained as

$$\delta = -\frac{2(m-1)(m+2)\left(\beta^2 t^2 + 2\alpha\beta t t_0 + (\alpha-1)\alpha t_0^2\right)}{3(2m+1)(\beta t + \alpha t_0)^2}. \tag{54}$$

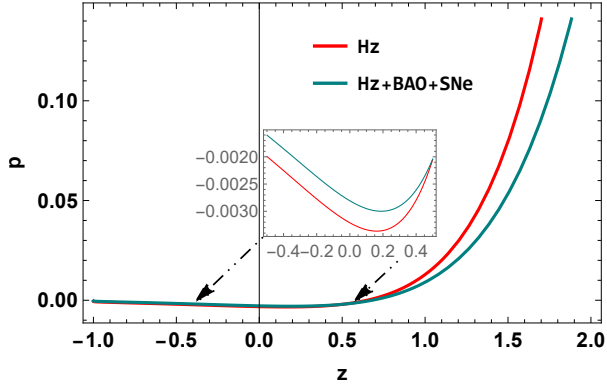

FIG. 7. Evolution of pressure $p$ in quadratic $f(Q)$ gravity model with the constraints value of the model parameters $\alpha$, $\beta$ together with $\lambda = -0.5$ and $m = 0.5$ versus redshift $z$.

From Fig. 7, it is clear that the anisotropic pressure is negative in the present and in the future. In the context of modified theories of gravity, negative pressure is caused by DE which is responsible for the acceleration of the current Universe. By definition, the skewness parameter is the quantity of anisotropy in a dark energy fluid, denoted by $\delta$. Fig. 8 represents its behavior in terms of redshift. We observed that that the skewness parameter is negative in the past ($z > 0$) and positive in the present ($z = 0$) and future ($z < 0$). Thus, our model is anisotropic throughout the expansion of the Universe.

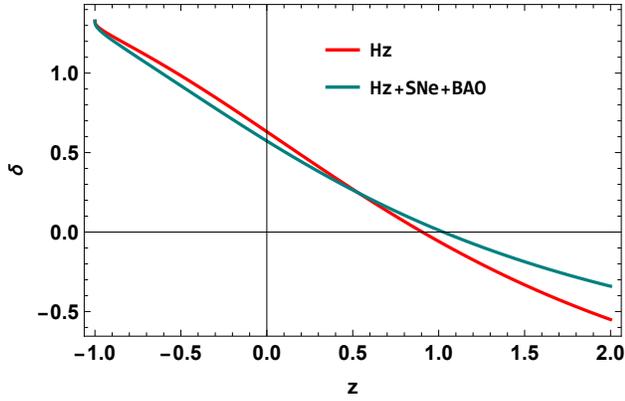

FIG. 8. Evolution of skewness parameter $\delta$ in quadratic $f(Q)$ gravity model with the constraints value of the model parameters $\alpha$, $\beta$ together with $\lambda = -0.5$ and $m = 0.5$ versus redshift $z$.

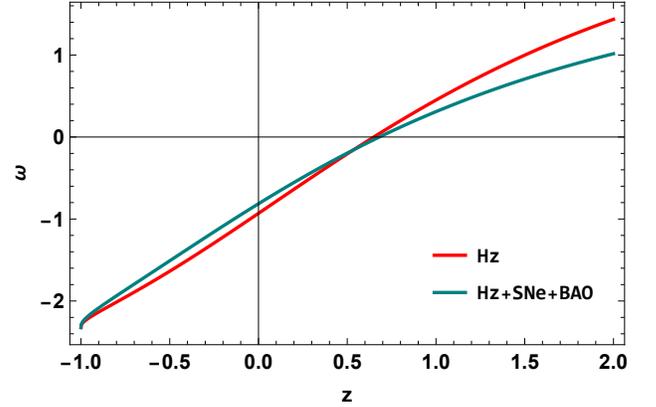

FIG. 9. Evolution of equation of state parameter $\omega$ in quadratic $f(Q)$ gravity model with the constraints value of the model parameters $\alpha$, $\beta$ together with $\lambda = -0.5$ and $m = 0.5$ versus redshift $z$.

Now, by using Eqs. (14) and (15), the skewness parameter is obtained as

Subtracting (6) from (7), we obtain the expressions of the EoS parameter in the model as

$$\omega = \frac{-\beta^2(2m+7)t^2 - 2\alpha\beta(2m+7)tt_0 + \alpha t_0^2(-7\alpha - 2(\alpha-2)m + 8)}{3(2m+1)(\beta t + \alpha t_0)^2} \tag{55}$$

From Eq. (30), we observe that the EoS parameter for our model is a function of redshift and approaches to $-1$ at present. It is known in the literature that the EoS parameter describes the evolution of the Universe from the early deceleration phase to the current acceleration phase. It includes stiff fluid epoch ($\omega = 1$), radiation epoch ($\omega = \frac{1}{3}$), dust epoch ($\omega = 0$), quintessence ($-1 < \omega < -\frac{1}{3}$), cosmological constant ($\omega = -1$) and the phantom ($\omega < -1$). Fig. 9 show the behavior of the EoS parameter with redshift $z$. It is observed that $\omega$ is situated in the quintessence region at present ($z = 0$) and over time $\omega < -1$ in the future (i.e. $z < 0$). Further, an interesting result of our model is that the current values of the EoS parameter i.e. $\omega_0 = -0.920$ for the Hubble dataset and $\omega_0 = -0.794$ for the $Hz$+SNe+BAO dataset, are in good agreement with recent Planck ob-servational data. In Ref. [12], Aghanim et al. give the constraints on the EoS parameter for DE. It can be seen that our estimates of the current value of the EoS parameter are within the observational limits mentioned in the previous reference. It is worth pointing out that the observations made in Ref. [29] are identical with our findings for the future time.

## VI. ENERGY CONDITIONS

In this section, we analyze the energy conditions (ECs) to validate the physical viability of our model. The ECs in $f(Q)$ gravity are given by [26]

- Weak energy conditions (WEC) if $\rho \geq 0$;
- Null energy condition (NEC) if $\rho + p \geq 0$;
- Dominant energy conditions (DEC) if $\rho - p \geq 0$;
- Strong energy conditions (SEC) if $\rho + 3p \geq 0$.

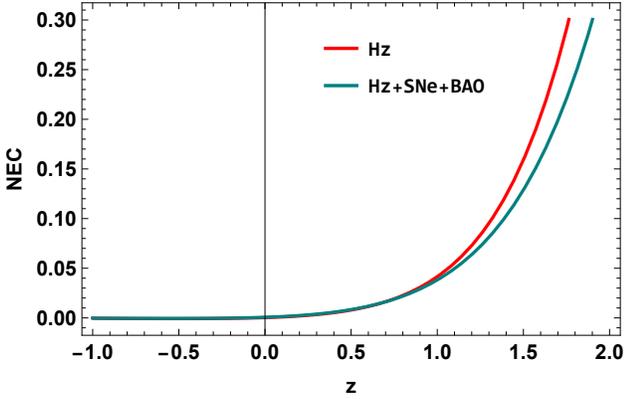

FIG. 10. Evolution of NEC in quadratic $f(Q)$ gravity model with the constraints value of the model parameters $\alpha$, $\beta$ together with $\lambda = -0.5$ and $m = 0.5$ versus redshift $z$.

In addition, the ECs in symmetric teleparallel $f(Q)$ gravity are discussed in Ref. [26]. In our analysis, we have considered the behaviour of these ECs, as displayed in Figs. 10-12. It is noted that $\rho + p \geq 0$, and $\rho - p \geq 0$, which represents that NEC and DEC, respectively are satisfied in the present and the future, while $\rho + 3p \geq 0$ (SEC) is violated. Hence, the violation of SEC leads to the acceleration of the Universe.

## VII. THERMODYNAMICAL ASPECTS AND ENTROPY OF THE UNIVERSE

In this section, we discuss the thermodynamical aspects of our analysis. According to the second law of thermodynamics, the entropy of the horizon (inside and the boundary) is always positive and increases with cosmic time. However, applying the combination of the first and second-law of thermodynamics to the system with volume $V$, leads to [55]

$$TdS = d(\rho V) + pdV = d\left[(\rho + p)V\right] - Vdp, \quad (56)$$

where $T$ and $S$ represent the temperature and entropy, respectively. In a thermodynamical system, the relation between the pressure (or energy density) and temperature for the perfect fluid is given by

$$dp = \left(\frac{p+\rho}{T}\right) dT. \quad (57)$$

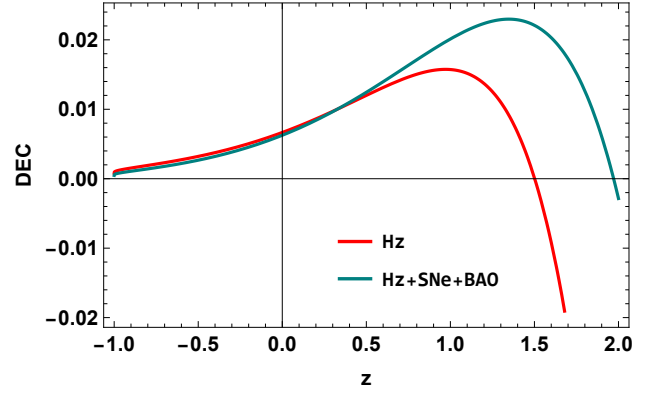

FIG. 11. Evolution of DEC in quadratic $f(Q)$ gravity model with the constraints value of the model parameters $\alpha$, $\beta$ together with $\lambda = -0.5$ and $m = 0.5$ versus redshift $z$.

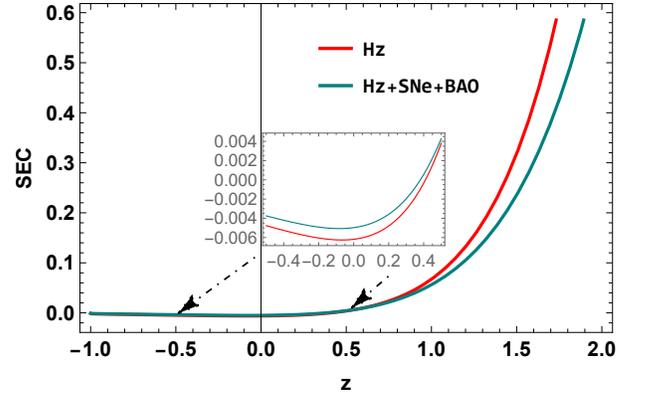

FIG. 12. Evolution of SEC in quadratic $f(Q)$ gravity model with the constraints value of the model parameters $\alpha$, $\beta$ together with $\lambda = -0.5$ and $m = 0.5$ versus redshift $z$.

By inserting Eq. (57) into Eq. (56), we have the differential relation

$$dS = \frac{1}{T} d\left[(p+\rho)V\right] - (p+\rho)V\frac{dT}{T^2} = d\left[\frac{(p+\rho)V}{T} + C\right], \quad (58)$$

where $C$ is a constant. Now by integrating the above equation, one can obtain the expression for entropy as follows

$$S = \frac{(p+\rho) V}{T}. \quad (59)$$

In addition, for an adiabatic process i.e. $S = const$, we can find the same definition of entropy starting from the conservation law which can be rewritten as $d\left[(\rho+p)V\right] = Vdp$. We denote the entropy density by $\widetilde{S}$ and rewrite Eq. (59) as

$$\widetilde{S} = \frac{S}{V} = \frac{(p+\rho)}{T} = \frac{(1+\gamma)\rho}{T}, \quad (60)$$

where $0 < \gamma < 1$. Now, to find the entropy density in terms of temperature, the first law of thermodynamics may be formulated as

$$d(\rho V) + \gamma \rho dV = (1+\gamma) T d\left(\frac{\rho V}{T}\right). \quad (61)$$

After simple calculations, from the above equations, we can get the expressions for entropy density and temperature as follows

$$\widetilde{S} = (1+\gamma)\rho^{\frac{1}{1+\gamma}}, \quad (62)$$

$$T = \rho^{\frac{\gamma}{1+\gamma}}. \quad (63)$$

Using Eq. (48) and (30) we find the entropy density and temperature for our analysis in the form

$$\widetilde{S} = (1+\gamma)\left[-\frac{486\lambda(2m+1)^2}{(m+2)^4}\left(\frac{\alpha}{t}+\frac{\beta}{t_0}\right)^4\right]^{\frac{1}{1+\gamma}}, \quad (64)$$

$$T = \left[-\frac{486\lambda(2m+1)^2}{(m+2)^4}\left(\frac{\alpha}{t}+\frac{\beta}{t_0}\right)^4\right]^{\frac{\gamma}{1+\gamma}}. \quad (65)$$

From Fig. 13, it is clear that the temperature for our model is an increasing function of redshift for the constrained values of the model parameters $\alpha$ and $\beta$. Further, the thermodynamic temperature is infinite in the early phase ($z > 0$) and with the cosmic evolution of the Universe, it decreases and finally attains a constant value in the future epoch ($z \to -1$). Hence, the model confirms the validation of second-law of thermodynamics. From Fig. 14, it appears that the entropy density of the Bianchi type-I cosmological model is a positive increasing function of redshift. It can be noticed that its behavior is totally similar to that of the energy density in Eq. (52). Moreover, the total entropy of the Universe increases during cosmological evolution i.e. $d\widetilde{S} \geq 0$ [56].

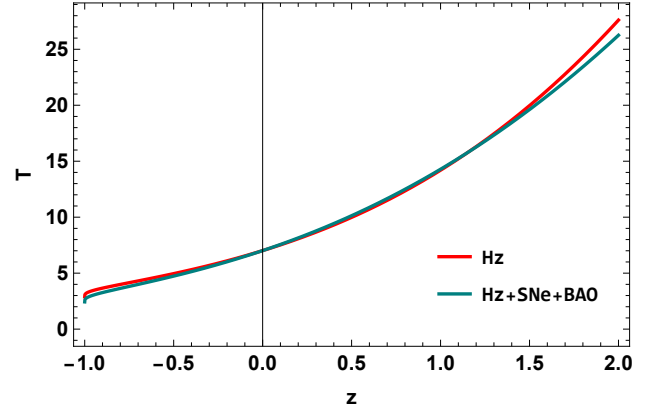

FIG. 13. Evolution of thermodynamic temperature $T$ in quadratic $f(Q)$ gravity model with the constraints value of the model parameters $\alpha$, $\beta$ together with $\lambda = -0.5$, $m = 0.5$ and $0 < \gamma < 1$ versus redshift $z$.

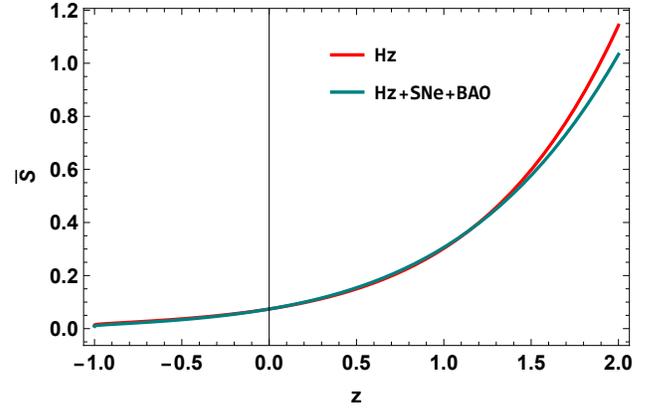

FIG. 14. Evolution of thermodynamic entropy density $\overline{S}$ in quadratic $f(Q)$ gravity model with the constraints value of the model parameters $\alpha$, $\beta$ together with $\lambda = -0.5$, $m = 0.5$ and $0 < \gamma < 1$ versus redshift $z$.

## VIII. COSMIC JERK PARAMETER

The cosmic jerk parameter is one of the basic physical quantities to explain the dynamics of the Universe. The Jerk parameter is a dimensionless third derivative of the scale factor $a(t)$ with respect to cosmic time $t$ and it is defined as [57]

$$j = \frac{\dddot{a}}{aH^3}. \quad (66)$$

Eq. (66) can be written in terms of a deceleration parameter $q$ as

$$j = q + 2q^2 - \frac{\dot{q}}{H}. \quad (67)$$

Using Eqs. (20) and (21), the jerk parameter for our

model is

$$j = 1 + \frac{(2t_0 - 3\beta t - 3\alpha t_0)\,\alpha t_0^2}{(\beta t + \alpha t_0)^3}. \tag{68}$$

The value of the jerk parameter is $j = 1$ for $\Lambda CDM$ model. The transition of the Universe from the decelerating to the accelerating phase results in a positive jerk parameter $j_0 > 0$ and a negative deceleration parameter $q_0 < 0$ corresponding to $\Lambda CDM$ model. From Fig. 15, we can see that the cosmic jerk parameter is positive throughout the evolution of the Universe and approaches 1 at late times. Thus, our model confirmed $\Lambda CDM$ model in the future.

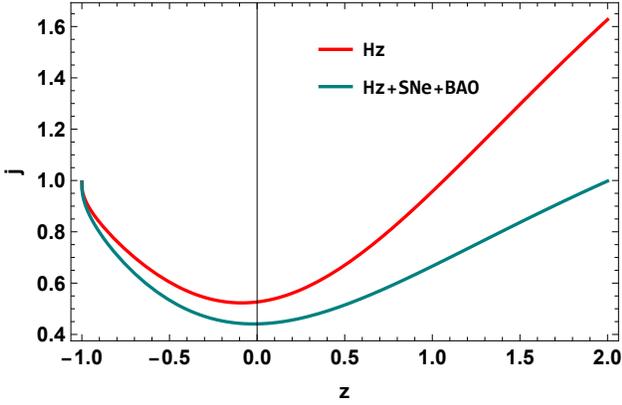

FIG. 15. Evolution of jerk parameter $j$ in quadratic $f(Q)$ gravity model with the constraints value of the model parameters $\alpha$, $\beta$ versus redshift $z$.

## IX. CONCLUSIONS

In this work, we investigated the Bianchi type-I cosmological model in the framework of $f(Q)$ gravity theory. We considered a quadratic $f(Q)$ model, specifically, $f(Q) = \lambda Q^2$, where $\lambda$ is the free model parameter. Then, we derived the field equations for the flat Bianchi type-I Universe and find the exact solutions using the relation between cosmic time and redshift as $t = \left(\frac{\alpha t_0}{\beta}\right) W\left[\frac{\beta}{\alpha} e^{\frac{\beta - \ln(1+z)}{\alpha}}\right]$. The features of the derived model are analyzed with constraints values of the model parameters $\alpha$ and $\beta$. The Hubble dataset, which consists of 57 data points, 1048 points from the SNe Ia dataset, and 6 points from the BAO dataset, was also utilized to constrain the model parameters. For the Hubble dataset and the combined $Hz$+SNe+BAO dataset, we have computed the best fit values of the model parameters. The model parameters that were found to suit the data the best are presented in Tab. II. Our findings are summarised as follows: The evolution of our model initiates with zero volume and an infinite rate of expansion which slows down for later epochs. The model incorporates anisotropy due to the radial and tangential stresses being unequal. Our model shows a transition from early deceleration to late-time acceleration and acquires a fixed value of the deceleration parameter which is in excellent agreement with current observations. According to the Hubble and $Hz$+SNe+BAO datasets, the present value of the deceleration parameter is $q_0 = -0.477$ and $q_0 = -0.427$, respectively. Further, the energy density for our model is positive and increasing function of redshift while the anisotropic pressure is negative at present and in the future. In addition, The evolution of the skewness parameter shows that our model is anisotropic throughout the expansion of the Universe. From the behavior of the EoS parameter, it is observed that $\omega$ is situated in the quintessence region at present and over time $\omega < -1$ in the future. Also, the current values of the EoS parameter i.e. $\omega_0 = -0.920$ for the Hubble dataset and $\omega_0 = -0.794$ for the $Hz$+SNe+BAO dataset, are in good agreement with recent Planck observational data. Furthermore, we have analyzed the behavior of different energy conditions, and we obtained that all the energy conditions are satisfied at present and the future while the SEC is violated. From a thermodynamic, we observed that the temperature for our model is an increasing function with respect to redshift. The thermodynamic temperature is infinite in the early phase and with the cosmic evolution of the Universe, it decreases and finally attains a constant value. The entropy density is a positive increasing function and its behavior is totally similar to the energy density. Moreover, the total entropy of the Universe increases during cosmological evolution. Finally, the cosmic jerk parameter is positive throughout the evolution of the Universe and approaches to 1 at late times. Thus, our model is in accordance with the recent observational data.

## ACKNOWLEDGMENTS

We are very much grateful to the honorary referee and the editor for the illuminating suggestions that have significantly improved our work in terms of research quality and presentation.

**Data availability** There are no new data associated with this article

**Declaration of competing interest** The authors declare that they have no known competing financial interests or personal relationships that could have appeared to influence the work reported in this paper.

---